\documentstyle[12pt]{book}
\begin{document}

\sloppy
\raggedbottom

\chapter* 
{HOW TO RUN ALGORITHMIC INFORMATION THEORY ON A COMPUTER\footnote
{\it
Lecture given Friday 7 April 1995 at the Santa Fe Institute, Santa
Fe, New Mexico.  The lecture was videotaped; this is an edited
transcript.
}
}
\markright
{How to Run Algorithmic Information Theory on a Computer}
\addcontentsline{toc}{chapter}
{How to run algorithmic information theory on a computer}

\section*{Complexity, to appear}
\section*{}

\section*{G. J. Chaitin,
IBM Research Division,
P.~O. Box 704, Yorktown Heights, NY 10598,
chaitin@watson.ibm.com}
\section*{}

Hi everybody!  It's a great pleasure for me to be back here at the
new, improved Santa Fe Institute in this spectacular location.  I
guess this is my fourth visit and it's always very stimulating, so I'm
always very happy to visit you guys.  I'd like to tell you what I've
been up to lately.  First of all, let me say what algorithmic
information theory is good for, before telling you about the new
version of it I've got.

In my opinion the most important application of this theory is to show
the limits of mathematical reasoning.  And in particular what I've
constructed and exhibited are mathematical facts which are true for no
reason.  These are mathematical facts which are true by accident.  And
since they're true for no reason you can never prove whether they're
true or not.  They're sort of accidental mathematical facts.  They're
mathematical facts which are analogous to a coin toss, because
independent tosses of a fair coin has got to come out heads or tails
but there's no reason why it should come out one or the other.  And
I've found mathematical facts which mirror this very precisely.  So
this is what algorithmic information theory is good for.

Now I've been working on this for a long time.  In fact for more than
thirty years---a life misspent!  There's basically three different
stages---well, one could find more stages in this theory---but there
are three main stages in the development of this theory.  There's the
old version of the theory which comes from the mid 1960's, and in that
version of the theory\ldots

But first of all I should say what algorithmic information theory is.
Basically it's recursive function theory plus one new idea which is
program-size complexity.  So the equation you want to have in your
mind is algorithmic information theory equals recursive function
theory plus one more notion which is to look at the size of programs,
that's the new element:
\[
   \mbox{AIT} = \mbox{recursive function theory} +
                \mbox{program size} .
\]
So it's roughly at the level of the kind of thing that Turing and
G\"odel were doing, but we throw in a complexity measure which is
program-size complexity.

Now this theory basically comes in three installments.  And I'll be
telling you about the third installment which I've created in the past
year.  But first let me tell you about the two previous installments
of this theory.  AIT$_1$, from the mid 1960's, is like this.  You look
at an $n$-bit string and you ask what is the size of the smallest
program for it, and it's usually about $n$ bits.  Most $n$-bit
strings require programs of about the same size they are.  So an
$n$-bit string usually has program-size complexity $n$.  That's the
first version:
\[
   \mbox{AIT$_1$: $n$-bit string, complexity $n$.}
\]

Then there's a new and improved---not new any more!---algorithmic
information theory from ten years later, the mid 1970's.  I like to
call that the self-delimiting version of the theory.  There the
basic idea is that you should be able to concatenate subroutines.
Information should be subadditive, it should be additive.  This sounds like
a technical detail but the whole theory is transformed when you make
this change.

Now, $n$-bit strings don't have complexity $n$.  Instead most $n$-bit
strings have what complexity?  Well, it's not just you need to know
what the $n$ bits are, you also need to know how many bits you're
getting.  So it's usually $n$ plus the base-two logarithm of $n$,
roughly speaking.  It's really $n$ plus the program-size complexity of
$n$, to give a more precise statement.
\[
   \mbox{AIT$_2$: $n$-bit string, complexity $n+\log_2n$.}
\]
So this is the idea that a program should be self-delimiting, should
indicate within itself its own size.  And I'm going to be talking more
about this.  Because I've found a new way of defining this.  One of
the problems with this theory is how to explain this.

Now what is the new algorithmic information theory which I view as
major rewrite three of the theory which I've just done in the past year?
One key idea I had in a sleepless night at your home, John [Casti], after
drinking too much and eating too much!  So the Santa Fe Institute is
involved with the genesis of this theory!  What this AIT$_3$ is, is it's
formally identical to algorithmic information theory$_2$, from an
abstract mathematical viewpoint there's no difference.  But there's
a world of difference!  Let me tell you what the world of difference is.
\[
   \mbox{AIT}_3  = \mbox{AIT}_2 + \Delta .
\]

The difference is basically two-fold.  Let's go back to recursive
function theory in the equation
\[
   \mbox{AIT} = \mbox{\bf recursive function theory} +
                \mbox{program size} .
\]
There's lots of books on recursive functions, like Hartley Rogers [1],
I noticed it in your library.

If you look at this book, let me tell you my reaction on looking at
this book.  This book is talking about computer programs, right?  And
all that Hartley Rogers cares about and all my theory cares about is
whether the program {\bf eventually} halts.  You don't care how long
it takes.  What I add is that I do care about the size in bits of the
program:
\[
   \mbox{AIT} = \mbox{recursive function theory} +
                \mbox{\bf program size} .
\]
But the interesting thing about these books from the point of view of
a computer programmer is that they're lousy books because they're
talking all the time about programs but there isn't one program that
you can actually run.  Maybe that was okay then, but I like actually
using computers!  So I'd like to tell you how to take algorithmic
information theory---which roughly speaking, is how to take recursive
function theory---and actually have it running on the computer.
Think of taking Hartley Rogers and writing out programs.  Basically
the way I do this is I use pure LISP.  But that's not enough.  Pure
LISP isn't good enough.  But basically speaking my position is that
the right way to do recursive function theory is to use pure LISP as
the programming language:
\[
   \mbox{AIT} = \mbox{recursive function theory ({\bf pure lisp})}
              + \mbox{program size} .
\]
But that's not good enough for algorithmic information theory.  Pure
LISP is good enough as is for recursive function theory.  So I'm going
to tell you what changes you need to make to pure LISP in order that
it should work for actually writing out programs in algorithmic
information theory.

Let me say more forcefully why this is important.  Algorithmic
information theory says your complexity measure is the size of the
smallest program for a universal Turing machine, and in AIT$_2$ it's
got to be self-delimiting.  Now there are two problems.  First of all,
which universal Turing machine do you pick?  They're all pretty much
equivalent but they give you slightly different complexity measures
depending on the choice of universal Turing machine.  Every universal
Turing machine can simulate every other, but having to tell it which
machine to simulate adds a certain number of bits to the complexity.

The other problem with the universal Turing machines used in
algorithmic information theory versions 1 and 2 is that you construct
them mathematically but they're not machines that you can actually use
to write programs and run them, and I think that they should be.

So I have two things I want to do.  On the one hand I want to make
algorithmic information theory, the theory of program-size complexity,
be the size of programs in an actual powerful, usable programming
language, instead of an abstract, theoretical programming language
which you can't use.  I'm going to tell you how to do that, and I'll
start with LISP.  And the other thing I want is to actually pick one
universal Turing machine, and let me emphasize why it's important
to do that.  During a visit here to the Santa Fe Institute I had a
conversation with Carlton Caves, who's a physicist.  Carl was interested
in applying program-size complexity to gedanken experiments in
thermodynamics and statistical mechanics like Maxwell's demon.
A typical thing that he would think about is you have a chamber divided in
half, and you have Avogadro's number of gas molecules.  He was interested
in whether each gas molecule is on the left or on the right in this
chamber divided in two.
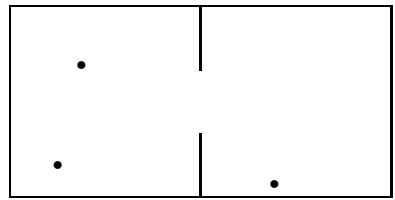
\begin{center}
\begin{picture}(144,72)(0,0)
\put(0,0){\framebox(144,72)}
\put(72,0){\line(0,1){24}}
\put(72,48){\line(0,1){24}}
\put(18,12){\circle*{3}}
\put(27,50){\circle*{3}}
\put(100,5){\circle*{3}}
\end{picture}
\end{center}

Is Avogadro's number about $10^{27}$?  Oh, it's about $10^{23}$?  So
you have about $10^{23}$ gas molecules and you get a bit string which
is $10^{23}$ bits long where a zero means that a particular molecule
is on the left and a one means it's on the right.  Then Carl would use
as his entropy measure the size of the smallest program that
calculates this $10^{23}$-bit string.  But there's a problem.  Since
this theory AIT depends on the choice of universal Turing machine, you
might pick a universal Turing machine which would completely swamp the
$10^{23}$ bits.  You want to know that the choice of the machine is
not going to make a difference which will be that big, otherwise my
theory AIT cannot be applied in this particular case.  So it'd be nice
to pick a particular machine and know what the complexity measure will
be.  So I've picked a particular machine, and that gives a much more
concrete version of the theory.  And later on I'm going to give you a very
concrete version of one of my incompleteness results.

But now let me tell you what we have to do to LISP so that you'll be
able to write programs in algorithmic information theory and how I pick
a particular universal Turing machine to use to measure the size of
programs.   It's not that difficult, it's a few simple ideas.  Here's
the first step.

Since in my view the main application of algorithmic information theory
is to study the limits of mathematical reasoning, I need some way to
talk about the complexity of a formal axiomatic system.  What's a
formal axiomatic system?  The abstract view is that it's a recursively
enumerable set of assertions.  In other words, you have a
proof-checking algorithm as Hilbert emphasized.  You run through all
possible proofs in size order, and you check which proofs are correct.
That means that given a set of axioms and a set of rules of inference
which are specified very, very precisely, you can in principle just
sit down and run through all possible proofs in size order and check
which ones are correct---it's like a million monkeys typing away.
This way in principle you can print out all the theorems.

So the abstract view I take is that a formal axiomatic system is just
a set of strings, a set of propositions, that there's some algorithm
for printing out in arbitrary order.  The point is that it's an infinite
computation.  A formal axiomatic system is an unending computation
that prints out an infinite set of strings which are the theorems---that's
the abstract point of view.  And my AIT gives you results limiting the
power of a formal axiomatic system in terms of its complexity.  What
is its complexity?  It's the size in bits of the smallest program that
will print out all the theorems.  So I need this complexity measure.

Now pure LISP, what is pure LISP like?  Those of you who are LISP
experts, please forgive me!  Well, roughly speaking, pure LISP is like
set theory.   In my opinion, pure LISP is to computational mathematics
as set theory is to theoretical, non-computational mathematics.
I don't know why more people don't realize this!  Maybe it's unusual
for people to simultaneously care about pure mathematics and actually
writing out programs.

Instead of having the set
\[
   \{ 1, 2, 3 \} ,
\]
which is unordered, what pure LISP has is a list
\[
   ( 1 \; 2 \; 3 )
\]
where you use parentheses and you don't use commas to separate things.
The main difference is that
\[
   ( 1 \; 2 \; 3 )
\]
is an ordered set (list!), in other words, there's a first, a second, and
a third element.  And you can nest things arbitrarily deep:
\[
   ((1) \; (2 \; 3)) .
\]
In pure LISP this is the data and these things are also the programs;
everything is constructed out of this.  It's the substance out of which
you build your universe.

Also LISP is a functional language, it's not an imperative language.
In other words, what you have are mathematical expressions for breaking
apart and putting together lists.  You don't {\bf do} anything, there's
no notion of time in pure LISP.  Instead you define functions and then
you apply the functions to arguments to see the values you get.
You don't have goto's, and you don't have assignment statements.  Instead it's
very much like mathematics in that you have expressions and you evaluate them.
So it's a functional language and a pure LISP program is actually a large
expression and you evaluate it and it gives you a value.

There's only one problem with this beautiful functional notion,
with this arithmetic of lists, not numbers, with this expression-based
language that's so clean and mathematical---I'm talking about pure LISP
with no side-effects!---there's only one problem with this which is how
do you compute infinite sets?  You can't!  But I want to be able to
print out one by one all the theorems of a formal axiomatic system!

It's very easy---here's how I add this to LISP.  First of all, you put
into LISP a primitive function called {\sc display}:
\[
   ( \mbox{display} \; \ldots ) .
\]
This is an identity function; the value of this
\[
   ( \mbox{display} \; \ldots )
\]
is the same as the value of the argument.  Oh by the way, in LISP
\[
   f(x,y)
\]
is written like this
\[
   (f \; x \; y) .
\]
That's the LISP notation for everything.
So this
\[
   ( \mbox{display} \; \ldots )
\]
is just the identity function, but it does have a side-effect, which is
to display the value of its argument.  This is actually used in normal
LISP for debugging.  It's a way to get more than the final value, it's
a way to look at intermediate results.

So here's how you use pure LISP to program out a formal axiomatic
system.  A formal axiomatic system is a LISP expression whose
evaluation will usually never finish.  But you don't care about the
final value, if any, what you care about are the intermediate values,
which are the theorems which you output using {\sc display}.  So it's
an evaluation which starts and will go on forever (but it might halt
if there are only a finite number of theorems), and each theorem is
put out as an intermediate result like this:
\[
   ( \mbox{display} \;\; \mbox{theorem} ) .
\]
This already exists in normal LISP.  But it's not enough.  If somebody
gives you a formal axiomatic system---that's a LISP expression whose
evaluation will never complete and which will put out one by one these
intermediate results---you need some way to get those intermediate
results.  So I add to LISP a primitive function called {\sc try}
that's very, very important.  {\sc try} has two arguments.  One is a
time limit and other is some LISP expression, which will often be a
formal axiomatic system:
\[
   ( \mbox{try} \;\; \mbox{time-limit} \;\;
     \mbox{formal-axiomatic-system} ) .
\]
What {\sc try} does is it starts the formal axiomatic system going,
and it runs it for a limited amount of time, and {\sc try} captures
the intermediate output, it captures the theorems.  {\sc try} is like
{\sc eval}.  There's a thing called {\sc eval} in LISP.  You can put
together an expression and then you can run it, and what {\sc eval}
does is it tells you the value that you get by running the expression.
{\sc try} is like a time limited {\sc eval}, plus it gives you the
intermediate results too.  You can't {\sc eval} a formal axiomatic
system, that's an infinite computation and you never get anything back
from {\sc eval}!  But you can {\sc try} a formal axiomatic system, and
then you get back three different things.  If the evaluation of the
LISP expression completes, {\sc try} will say that it completed and
will give you its value.  If not, {\sc try} will let you know that it
ran out of time.  And in either case {\sc try} will let you know all
the intermediate results, all the theorems, that were displayed.  So
the value that you get back from {\sc try} is a pair:
\[
   ( \mbox{value/out-of-time} \;\; \mbox{captured-displays} ) .
\]
If the expression being tried is a formal axiomatic system with an
infinite number of theorems, then the first element of the pair coming
back from a {\sc try} will always say that it ran out of time, and the
second element of the pair will be all the theorems that it managed to
prove before the time limit ran out.

Okay, so this gives you a way to deal with infinite sets in LISP.
Normal LISP cannot deal with unending computations.  But that's not
enough.  We also have to add binary data to LISP.  Why?

The obvious way to get a program-size complexity measure using LISP is
to use as your measure the size in characters of LISP expressions,
which are actually called ``S-expressions'' in LISP.  This is a nice
concrete program-size measure, but it doesn't give you the correct
complexity measure of AIT$_2$.  What we need is a way to give LISP
raw binary data, because LISP expressions aren't a good way to package
algorithmic information because LISP syntax means that there's
redundancy and you're not using all the bits efficiently enough.  So
what we really want is a LISP expression plus a way to give it raw
bits on the side.  What does it mean for a LISP expression to have
access to binary data ``on the side?''  It means that now the
environment in which a LISP expression is evaluated doesn't just
include the current variable bindings, it also includes a list of
bits, a list of zeros and ones.  How does a LISP expression get access
to this binary data?  Well, we provide two new primitive functions for
doing this, one to read the next bit, and one to read a complete LISP
expression from the binary data.  These are functions with no explicit
arguments: you just write
\[
   ( \mbox{read-next-bit} )
\]
or
\[
   ( \mbox{read-next-S-expression} ) .
\]
In the first case, {\sc read-next-bit}, what you get when you evaluate
it is either a zero or a one.  It's the next bit of the binary data,
{\bf if there is a next bit!} It's very important that if you've used
up all the binary data and there is no next bit to read, then {\sc
read-next-bit} explodes.  If you've run off the end of your binary
data, then {\sc read-next-bit} fails, which is very important, as I'll
explain later.  In fact, this is the key step in getting AIT$_2$ out
of LISP, that if you try to read a bit that isn't there you explode.
What about {\sc read-next-s-expression}?  What it does is it reads, say,
eight bits at a time and interprets them as a character in a LISP
expression.  It keeps reading until parentheses balance and it has a
complete LISP expression, or until it runs out of bits and fails.
Actually, it's seven bits per ASCII character that it reads.

The next question is, how do you give binary data to a LISP expression
that wants some raw bits on the side?  Well, you do it with {\sc try},
I'm making {\sc try} work overtime!  {\sc try} actually has three
arguments.  There's a time limit, there's an expression to be
evaluated, and finally there's the binary data.  So {\sc try} ends up
having three arguments:
\[
   ( \mbox{try} \;\; \mbox{time-limit} \;\;
     \mbox{expression} \;\; \mbox{binary-data} ) .
\]
{\sc try} is really at the heart of my whole new theory AIT$_3$!  If
you understand this you understand everything that I've added to LISP
to get algorithmic information theory to work.  We've already seen the
time limit and the expression that you try to evaluate for that amount
of time.  What's new is the third argument, the binary data, which
could in fact be the empty list, in which case there is no binary
data.  And what's also new, and this is very, very important, is that
now a {\sc try} can fail in two ways!  It can fail because you run out
of time.  Or it can fail because you run out of binary data, because
you tried to read bits that weren't there.  So the value that {\sc
try} returns now looks like this:
\[
   ( \mbox{value/out-of-time/out-of-data} \;\;
     \mbox{captured-displays} ) .
\]

{\sc Question:} How do you measure how much time it takes the
expression to execute?

{\sc Answer:} That's a good question but I should answer it in
private---it's a mess!

{\sc Question:}  Is it related to how many evaluations you have to do?

{\sc Answer:} There are many possibilities.  Actually I use the
interpreter stack depth as my time limit, but there are many
possibilities and I'm not sure I picked the right one!

Okay, so this is how we give binary data to a LISP expression:
\[
   ( \mbox{try} \;\; \mbox{time-limit} \;\;
     \mbox{expression} \;\; \mbox{binary-data} ) .
\]
And it can fail either because it runs out of time or because it runs
out of data.  Now what is my program-size complexity measure?  (What
I'm really going to tell you now is what's the universal Turing
machine I'm picking to measure program-size complexity with.)  Well
it's very simple!  I don't just have pure LISP any more; I've added
binary data to pure LISP.  So how do I measure the size of a program
now?  Well a program isn't just a LISP expression any more, because it
can have binary data on the side:
\[
   ( \mbox{try} \;\; \mbox{time-limit} \;\;
     \mbox{expression} \;\; \mbox{binary-data} ) .
\]
So there are now two parts to the program, the expression and the
data.  I take the LISP expression and I measure it's size in
characters.  Then I multiply by eight or seven bits per character.  Or
perhaps it's sixteen bits per character if you're using extended
characters for Japanese.  This gives me the size of the expression
measured in bits instead of characters.  And finally I just add the
number of bits in the binary data.  This is how I measure the size of
a LISP expression with binary data on the side, and this includes the
possibility that there's actually no binary data.  So LISP expressions
can now use two new primitive functions with no arguments to read a
single bit or a LISP expression from the binary data, and if they do
this they are charged one bit for each bit of the binary data that
they read.  So that's how I measure the size of a program now.  And
the program-size complexity of an object, of a LISP expression, is
defined to be the size of the smallest program, of the smallest
expression/binary data pair, that produces it, that yields it as its
value.

So this is how we give raw bits to LISP expressions.  And it is very
important to note that you fail if you run out of binary data.  You do
not get a graceful end-of-file indication!  If you did, we would get
AIT$_1$ out of LISP, not AIT$_2$ with self-delimiting programs.  And
why are our programs self-delimiting?  The LISP expression part is
self-delimiting because parentheses have to balance.  And the binary
data that the LISP expression reads is self-delimiting because we are
not allowed to run off the end of the data.  It follows that our
program-size complexity measure is additive.  This means that the
program-size complexity $H(x,y)$ of a pair of LISP expressions is
bounded by the sum of the individual complexities $H(x)$ and $H(y)$
plus a constant:
\[
   H(x,y) \le H(x) + H(y) + c .
\]
This only works with self-delimiting programs.  It does not work in
the original algorithmic information theory from the 1960's.

Let me explain another way what this complexity measure is.  Here is
how to reformulate what I've just explained using {\sc try}, another
way, using a universal Turing machine with binary programs.  In this
way of looking at it, I'm not really using LISP as my programming
language.  Instead this LISP is sort of a high-level assembler to
produce binary programs that I feed to a universal Turing machine.
This universal Turing machine reads its program from the binary data,
bit by bit.  The first thing it does is to read a complete LISP
expression from the beginning of the binary data, which just means
that it goes on until the parentheses balance.  Then the Turing
machine starts to run this prefix, to evaluate it, running it against
the remainder of the binary data (if any's left).  So there is a
prefix, which is read eight or seven or sixteen bits at a time, and
then the prefix starts to run and it can read in additional bits if it
wants to by using {\sc read-next-bit} or {\sc read-next-s-expression}.
And the prefix has to decide by itself how many bits to read, because
it's not allowed to discover that no bits are left.  If the prefix
asks for a bit that isn't there, then the whole thing fails, and this
wasn't a valid program for our universal Turing machine.

That turns out to be the whole story!  That's how to get algorithmic
information theory, and the right version of it, AIT$_2$, running on a
computer.  You see, it isn't hard to do if you like LISP programming!
I should say one thing though.  This only works because computers are
so powerful now.  If I had had this idea years ago, I wouldn't have
been able to run any interesting examples, because the machines were
too small and too slow.

So now I've picked out a particular universal Turing machine and my
program-size complexity measure is very concrete, and I can actually
write out the programs in LISP.  Now let me tell you some of the sharp
results that I get in this new more concrete theory, AIT$_3$.

{\sc Question:} What is the relationship between LISP and the Turing
machine?

{\sc Answer:} Well, the best way to think about it is that I've used
LISP to write a simulator for my universal Turing machine.  But this
universal Turing machine doesn't use LISP as its language, it uses a
strange binary language in which the beginning of a program is the bit
string for a LISP expression that tells us how to get the remaining
bits of the program and what to do with them.  So the language I'm
really using isn't LISP.  I'm using LISP as a high-level assembly
language to create these bits strings and concatenate them.  To do
this I have to add another new primitive function to LISP, one which
converts a LISP expression into a bit string.  And I'm also using LISP
to write a simulator for my universal Turing machine.  That's a very
simple LISP program to write using {\sc try}.  In other words, I take
pure LISP and I add some stuff to it.  Then I use it like this: On
the one hand to define a universal Turing machine that runs binary
programs.  On the the other hand I use this augmented LISP to put
together the long binary programs that I feed to this universal Turing
machine.  So this universal Turing machine is programmed in LISP, but
its programs are not in LISP.

Okay, I think that by now you should get the idea how AIT$_3$ works.
So let me tell you what kind of result you get using this new
approach.  Algorithmic information theory now becomes very concrete.
Every time you have a theorem about program-size complexity, you can
now actually write down the program that proves the theorem, and the
size of this program gives you a precise numerical value for what was
previously an undetermined constant in the statement of the theorem.
Here is an important example, the inequality that
\[
   H(x,y) \le H(x) + H(y) + c .
\]
Let's go back to thinking about programs in the form of expression/data pairs:
\[
   ( \mbox{try} \;\; \mbox{time-limit} \;\;
     \mbox{expression} \;\; \mbox{binary-data} ) .
\]
So we have an expression/data pair that calculates a LISP expression
$x$, and another expression/data pair that calculates a LISP
expression $y$.  The above inequality states that you can combine them
to get an expression/data pair that calculates the list $(x\;y)$, and this
combined expression/data pair is exactly $c$ bits bigger than the sum
of the sizes of the given expression/data pairs.  The LISP programming
required to show this is trivial---although the programming details
would require some explanation.  It finally turns out that $c$ is
twenty characters which at seven bits per characters is exactly 140
bits.  So that's the value of the constant $c$:
\[
   H(x,y) \le H(x) + H(y) + 140 .
\]

Now this may not sound terribly exciting, but for me it was a tremendous
revelation!  Why?  Because I've been proving theorems about program-size
complexity all my life, but I never actually had a program in front of
me and I never actually measured its size, and I never knew what the
constant was in this inequality!  For all I knew, this constant could
have been $10^{99}$!  Now I know that it's only 140, thank goodness!
This is important for Carlton Caves, because if this constant were
large compared to Avogadro's number, then Carl couldn't use this basic
inequality in his program-size complexity analysis of Maxwell's demon.

Now let me tell you about my main incompleteness theorem and how it
looks in this new more concrete formulation of my theory.  My main
incompleteness result has to do with a number I call $\Omega$, which
is the halting probability of our universal Turing machine.  One of
the reasons that you want binary programs to be self-delimiting, to
indicate within themselves where they end, is so that you can define
this halting probability.  Here's how it works.  You take our
universal Turing machine, and each time it asks for a bit, feed it the
result of an independent toss of a fair coin.  This works because the
machine decides by itself how many bits to read.  That's why we can
define the probability that a program of any size will halt.  If
programs weren't self-delimiting, then there wouldn't be a natural
probability measure to put on the space of all programs.  If programs
weren't self-delimiting, then all the $n$-bit programs, if you give
each of them probability $2^{-n}$, would add up to probability unity,
and how do you throw in programs of different sizes?  So to be able to
have a halting probability defined over programs of {\bf any size},
these programs have to be self-delimiting.  So it's very important for
the universal Turing machine to decide by itself how many bits to
read.  Since it does, we get this halting probability $\Omega$ which
is a real number between zero and one.  It's the halting probability
of the specific universal Turing machine that AIT$_3$ is based on.
I explained before how this Turing machine works.  Since this is a
specific Turing machine, its halting probability $\Omega$ is now a
specific real number.  Before $\Omega$ depended on our choice of
universal Turing machine.  The same theorems applied to each of these
$\Omega$'s, but now its a specific $\Omega$ that we're thinking about.

My main result about $\Omega$, and about this particular $\Omega$ too,
is that $\Omega$ shows that you have randomness in pure mathematics.
Why?  Let's say that you're trying to use formal reasoning, you're
trying to use a formal axiomatic system to prove what the bits of
this halting probability are.  But you can't because these bits are
accidental, there's no reason why they should be what they are,
they're irreducible mathematical information.  Essentially the only
way to prove what an individual bit in a particular place in the binary
expansion of $\Omega$ is, whether it's a zero or a one, is to add that
fact as a new axiom.   In other words, each bit of $\Omega$ has got to
come out zero or one, but it's so delicately balanced whether it should
come out one way or the other, that we're never going to know.

That's my old result from AIT$_2$, but what is the new, more concrete
version of it that I get in AIT$_3$?  The new, concrete incompleteness
result is this: To determine $n$ bits of $\Omega$, you need a theory
of complexity at least $n-7581$.
For the first 7581 bits of the halting probability, it
might be that a formal axiomatic system can prove what these first
7581 bits are.  But afterwards, every time you want to prove what one
more bit of $\Omega$ is, you have to add a bit to the complexity of
the formal axiomatic system that you're using.

By the way, seven thousand bits is only a thousand characters which
is only twenty lines of LISP code.  In other words, my proof of this
incompleteness result involves only twenty lines of LISP code, if
you compress out all the blanks and the comments!

In other words, after the first 7581 bits of $\Omega$, every additional
bit is going to cost you!  In fact essentially the only way to be able
to get out of a formal axiomatic system a theorem telling you what that bit is,
is if you put the theorem in as a hypothesis, as a new axiom!  That means
that at that point reasoning is not really getting you anywhere any more.

{\sc Question:}  But up to that point?

{\sc Answer:}  Up to that point, you might just be able to do it all.

{\sc Question:}  Up to 7581 bits?

{\sc Answer:}  Yeah.

In fact, the first seven bits of this particular halting probability
$\Omega$ are all ones.  I'm telling you that it's impossible to know
the bits of the halting probability, but in fact I do know the first
seven bits!  This is an embarrassing fact, but now I know how bad it
can be.  Somewhere between the first seven bits and the first seven
thousand it becomes impossible!  The first seven bits are all ones,
but now I know that you can go out at most a thousand times more than
that.  After that, every time you want to prove what another bit of
the halting probability is, you have to add a bit to your axioms, you
have to add a bit to the complexity of your formal axiomatic system.

What exactly is the complexity of a formal axiomatic system?  Well,
the formal axiomatic system is now considered to be a LISP expression
with binary data on the side.  The formal axiomatic system is a
program that goes on forever printing out the theorems.  And you
measure the complexity of the formal axiomatic system by taking the
LISP expression, converting its size from characters to bits, and
adding that to the number of bits in the binary data.  So we're using
the same complexity measure for infinite computations that we do for
finite computations.  Yes?

{\sc Question:}  Just to be clear on a point, that number 7581, that's
dependent on the specific formal axiomatic system that you've chosen?

{\sc Answer:}  No, no, it's dependent on the particular universal
Turing machine that I've chosen!

{\sc Question:} The universal Turing machine?

{\sc Answer:}  That's right.

{\sc Question:}  So if you change the Turing machine then that number
changes?

{\sc Answer:} And the halting probability $\Omega$ changes depending
on the universal Turing machine.  But for this specific universal
Turing machine, I finally know what the number in my incompleteness
theorem is, it's 7581.

{\sc Question:}  You have a procedure for determining what the next bit
is that you have to add to the axiomatic system?

{\sc Answer:} No, I have a proof showing that if you want to get
another bit, then you've got to add a bit to the formal axiomatic
system.  At that point $\Omega$ becomes unknowable, because the only
way you're going to get additional bits of $\Omega$ from your theory
is if you put them in as new hypotheses, as new postulates.  That's
the point, $\Omega$ is irreducible mathematical information.  So it's
impossible to know more bits of $\Omega$.

{\sc Question:}  Are these bits of $\Omega$ arbitrary?

{\sc Answer:}  No, they're not arbitrary!  Let me tell you why.

Here's another thing I can do with my augmented LISP.  I have a
program that's only about ten lines of LISP which can actually compute
lower bounds on the halting probability.  Given $n$, this program
looks at all $n$-bit programs, runs them for time $n$, and divides the
number that halt by $2^n$.  That gives a lower bound on the halting
probability, and the lower bounds get better and better as $n$ gets
larger.  I have this program, and I've actually run it for all $n$ up
to 22.  I've looked at all 22-bit programs.  It's very easy to write
out this program!  This could never be done before.  So I find it very
exciting that I can actually write down a program that computes better
and better lower bounds on the halting probability.  In fact it gives
the halting probability in the limit from below.  So if I can write a
program to compute it in the limit from below, it seems to me like a
pretty definite number!

By the way, this most definitely doesn't mean that $\Omega$ is a
computable real number like $\pi$.  It isn't, because the convergence
of this thing is unbelievably slow.  You never know how far out to go
to get a given degree of accuracy.  If the halting probability were
computable, then it would be very easy to prove what its bits are.  It
would be like $\pi$, you'd just calculate them!  But at least I can
write out this program that computes lower bounds, and I can even run
it for small $n$.

So what have I done with all of this new stuff?  Where can you learn
more about my software, and how can you get your hand on it?  Well
what I did is I put together a hands-on, computer-oriented course that
I call ``The limits of mathematics.''  There are several versions of
this.  And you can get all this stuff from chao-dyn at xyz.lanl.gov.
What you do is you just go to Web address http://xyz.lanl.gov/.  From
there it's easy to find my stuff and get it downloaded.

The first version of this stuff that I sent to xyz.lanl.gov is
a preliminary version of all of this in which the LISP interpreter is
written in Mathematica [2].  You can also get this version of my
course from MathSource via http://www.wri.com/.  The LISP interpreter
is a few pages of Mathematica code.

Then a friend of mine, George Markowsky, he's a professor at the
University of Maine in Orono, and he invited me to give an intensive
short course on the limits of mathematics using this hands-on
approach.  George is unusual in that besides being a good
mathematician, he's extremely good with computers.  So with his help I
took this Mathematica program, and I converted it into C, so that it
could run on small personal computers.  The C version of my LISP
interpreter does exactly the same thing that the Mathematica version
does, but it's much faster.  It's also a much larger program, and the
code is much more difficult to understand.

So giving the course in Maine led to another version of it [3] with new
software.  Besides the LISP interpreter in C, I also improved the
LISP programs.  This second version of ``The limits of mathematics''
starts off with a reference manual for my LISP, and the rest of the
book is just software written in this LISP.  Each of the LISP programs
has comments.  The comments in the program tell you what theorem the
program proves.  And there's usually a constant in the statement of
the theorem, and that constant is the size of the program.  This book
may be the ultimate in constructive mathematics, but it is not easy to
read!

There's also an extended abstract, that summarizes it all in a few pages [4].

I think that with powerful modern technology like Mathematica and UNIX
workstations, if you talk about an algorithm in a mathematics paper,
then you should actually give the program in the paper.  If we could
all pretty much agree on a very high-level language like Mathematica,
then one could always include understandable programs in a mathematics
paper.  And the mathematics paper should be available over the Net,
because you really don't want it on paper, you want to be able to grab
the programs and run them!

So this is what I've done with algorithmic information theory and its
incompleteness theorems.  But the problem is that the result is a
reference manual and a lot of software, and it is not easy to understand.
In fact, there is a more aggressive version [5] of my course with much
smaller constants, that's even harder to understand!

If any of you want an excuse to visit the Black Sea, this summer there's
going to be a meeting in Romania where I'll give this course again and
I'll try to explain it better [6].

Maybe the problem is that my LISP is a bit Spartan.  It only allows
one-characters variable names, and arithmetic has to be programmed
out, it is not provided built in.  So perhaps I should take the
trouble to flesh out my LISP and make it friendlier [7].

And maybe I can encourage someone who is a good teacher to write a really
understandable book-length treatment of all of this, because I guess I
prefer doing research to writing text books!

One more thing.  I went to your library and I gave your librarian a
present, a book called {\it Nature's Imagination\/} which has an
article of mine on ``Randomness in arithmetic and the decline and fall
of reductionism in pure mathematics'' [8].  That's a talk I gave at
Cambridge University two years ago, and at that time I thought it
summarized everything fundamental that I had to say about the limits
of mathematics.  Then I came up with this new version of my theory,
AIT$_3$!  So AIT$_3$ is not in there, but it's a pretty understandable
summary of what I think all of this implies about how we should
actually do mathematics.  I gave a copy of this book to your library,
it's {\it Nature's Imagination\/} edited by John Cornwell.  My
article is also in a book [9] edited by two of you in the audience,
Anders Karlqvist and John Casti.  That also doesn't seem to be in your
library, and neither are my three books.  So when I get back to my
institution, IBM Research, in May, I'll send your librarian copies of
my three books [10,11,12].

There's also a tutorial article of mine [13] that will come out in the
first issue of your magazine {\it Complexity.}

{\sc Question:}  That book you're referring to by Anders and me, it's
definitely there.  It may not be on the shelf at the moment, but it's
in the library.

{\sc Answer:}  Okay, I'm glad it's there.  Well I'm going to send you
my other books, so they'll show up in your library.

Okay, that's basically it, unless there are comments, criticisms,
questions, or anyone wants to throw tomatoes?!  Yes, sir?

{\sc Question:}  What's the difference between the word delimiting
and limiting?

{\sc Answer:}  Self-limiting?  I don't know, in computer programming
languages people talk about delimiters.  And that's why I call them
self-delimiting programs.  Anyone want to suggest a better word for
this?  That's the best I could come up with!  Do you think that
``self-limiting'' programs is better than ``self-delimiting?''

{\sc Question:}  No, what is meant by self-delimiting?

{\sc Answer:} Well, the programs in the oldest version, the 1960's
version of algorithmic information theory, I call them blank-endmarker
programs.  That's because the program has 0's and 1's and then there's
a blank at the end and you can read the blank and realize that the
program is finished and there are no more bits.  That blank is a
delimiter.  Now I throw out the blank delimiter and say you're not
allowed to read the blank at the end.  That program has failed, it
explodes, it ran off the tape, the machine collapses in a heap!  So
that could be called a no-blank-endmarker program, but I call it a
self-delimiting program instead.  If you can come up with a better
name, please do!

Any other questions?  Okay, thank you very much!

\section*{References}

\begin{itemize}

\item[{[1]}]
H. Rogers, {\it Theory of Recursive Functions and Effective Computability,}
McGraw-Hill, 1967.

\item[{[2]}]
G. J. Chaitin,
``The limits of mathematics---course outline \& software,''
chao-dyn/9312006,
127 pp., December 1993.

\item[{[3]}]
G. J. Chaitin,
``The limits of mathematics,''
chao-dyn/9407003,
270 pp., July 1994.

\item[{[4]}]
G. J. Chaitin,
``The limits of mathematics---extended abstract,''
chao-dyn/9407010,
7 pp., July 1994.

\item[{[5]}]
G. J. Chaitin,
``The limits of mathematics IV,''
chao-dyn/9407009,
231 pp., July 1994.

\item[{[6]}]
G. J. Chaitin,
``A new version of algorithmic information theory,''
chao-dyn/9506003,
12 pp., June 1995,
{\it Complexity,} to appear.

\item[{[7]}]
G. J. Chaitin,
``The limits of mathematics---tutorial version,''
chao-dyn/9509010,
143 pp., September 1995.

\item[{[8]}]
G. J. Chaitin,
``Randomness in arithmetic and the decline and fall of reductionism in
pure mathematics,''
chao-dyn/9304002,
in J. Cornwell, {\it Nature's Imagination,}
Oxford University Press, 1995, pp.\ 27--44.

\item[{[9]}]
J. L. Casti and A. Karlqvist,
{\it Cooperation and Conflict in General Evolutionary Processes,}
Wiley, 1995.

\item[{[10]}]
G. J. Chaitin,
{\it Algorithmic Information Theory,}
revised third printing,
Cambridge University Press, 1990.

\item[{[11]}]
G. J. Chaitin,
{\it Information, Randomness \& Incompleteness,}
second edition,
World Scientific, 1990.

\item[{[12]}]
G. J. Chaitin,
{\it Information-Theoretic Incompleteness,}
World Scientific, 1992.

\item[{[13]}]
G. J. Chaitin,
``The Berry paradox,''
chao-dyn/9406002,
{\it Complexity\/} 1 (1995), pp.\ 26--30.

\end{itemize}

\end{document}